\documentclass[conference]{IEEEtran}
\IEEEoverridecommandlockouts
\usepackage{cite}
\usepackage{amsmath,amssymb,amsfonts}
\usepackage{algorithmic}
\usepackage{graphicx}
\usepackage{textcomp}
\usepackage{xcolor}
\usepackage{natbib}
\usepackage{subcaption}
\def\BibTeX{{\rm B\kern-.05em{\sc i\kern-.025em b}\kern-.08em
    T\kern-.1667em\lower.7ex\hbox{E}\kern-.125emX}}
\begin{document}

\title{Temporal Narrative Monitoring in Dynamic Information Environments}

\author{\IEEEauthorblockN{David Farr} 
\IEEEauthorblockA{\textit{School of Information Science} \\
\textit{University of Washington}\\
Seattle, USA \\
dtfarr@uw.edu}
\and
\IEEEauthorblockN{Stephen Prochaska}
\IEEEauthorblockA{\textit{School of Information Science} \\
\textit{University of Washington}\\
Seattle, USA \\
sprochas@uw.edu
}
\and
\IEEEauthorblockN{Jack Moody}
\IEEEauthorblockA{\textit{Whiting School of Engineering} \\
\textit{Johns Hopkins University}\\
Baltimore, USA \\
jmoody11@jhu.edu
}
\and
\IEEEauthorblockN{Lynnette Hui Xian Ng}
\IEEEauthorblockA{\textit{School of Computer Science} \\
\textit{Carnegie Mellon University}\\
Pittsburgh, USA \\
lynnetteng@cmu.edu}

\and
\IEEEauthorblockN{Iain Cruickshank}
\IEEEauthorblockA{\textit{School of Computer Science} \\
\textit{Carnegie Mellon University}\\
Pittsburgh, USA \\
icruicks@andrew.cmu.edu}
\and
\IEEEauthorblockN{Kate Starbird}
\IEEEauthorblockA{\textit{Human Centered Design Engineering} \\
\textit{University of Washington}\\
Seattle, USA \\
kstarbi@uw.edu}
\and
\IEEEauthorblockN{Jevin West}
\IEEEauthorblockA{\textit{School of Information Science} \\
\textit{University of Washington}\\
Seattle, USA\\
jevinw@uw.edu}
}

\maketitle

\begin{figure*}[t] 
\centering 
\includegraphics[scale=.5]{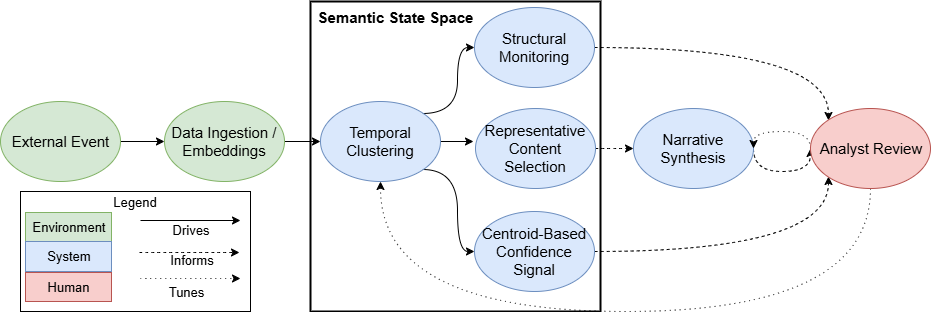} 
\caption{System architecture for temporal narrative monitoring. External events drive ingestion into a shared semantic space where temporal clustering produces evolving narrative clusters in a semantic state space. Centroid-based monitoring and generative summaries support analyst review and adaptive feedback.}
\label{sys_over} 
\end{figure*}

\begin{abstract}
Comprehending the information environment (IE) during crisis events is challenging due to the rapid change and abstract nature of the domain. Many approaches focus on snapshots via classification methods or network approaches to describe the IE in crisis, ignoring the temporal nature of how information changed over time.

This work presents a system-oriented framework for modeling emerging narratives as temporally evolving semantic structures without requiring prior label specification. By integrating semantic embeddings, density-based clustering, and rolling temporal linkage, the framework represents narratives as persistent yet adaptive entities within a shared semantic space. We apply the methodology to a real-world crisis event and evaluate system behavior through stratified cluster validation and temporal lifecycle analysis. Results demonstrate high cluster coherence and reveal heterogeneous narrative lifecycles characterized by both transient fragments and stable narrative anchors.

We ground our approach in situational awareness theory, supporting perception and comprehension of the IE by transforming unstructured social media streams into interpretable, temporally structured representations. The resulting system provides a methodology for monitoring and decision support in dynamic information environments.
\end{abstract}


\section{Introduction}
Understanding crisis information environments is challenging because online discourse changes rapidly as narratives form, compete, and dissipate in response to unfolding events. Existing analytical approaches commonly rely on classification-based NLP pipelines or network-based analyses of user interactions~\cite{ziems2024can, beers2023followback}. While effective in relatively stable settings, these methods struggle in dynamic environments where narratives evolve quickly, producing continual distributional change in both content and discourse patterns~\cite{gama2014survey, ng2022crossplatform}.

Classification workflows typically require predefined labels, implicitly assuming stable categories and prior knowledge of relevant narratives~\cite{ziems2024can}. Network-based approaches instead emphasize interaction structure, providing insight into information flow but limited visibility into how narrative meaning itself evolves over time~\cite{starbird2019disinformation}. Together, these limitations motivate treating narratives not as static labels or latent topics, but as evolving system-level entities.

In this work, we present a system-oriented framework that models narratives as temporally evolving semantic structures without requiring prior label specification. We make three contributions: (1) a scalable NLP system for detecting and tracking narratives in non-stationary social media streams, (2) a clustering-based temporal modeling approach enabling measurement of narrative emergence and drift, and (3) an empirical evaluation demonstrating system behavior during a real-world crisis event in Venezuela. Unlike prior topic modeling or clustering approaches that produce independent temporal snapshots, the proposed framework treats narrative structure as a persistent system state evolving within a shared semantic space. This enables measurement of narrative continuity, drift, and stability as observable system properties, transforming discourse monitoring from static analysis into dynamic system monitoring.

\section{Background and Related Work}

This work integrates concepts from multiple research areas, including situational awareness, natural language processing, and computational social science.  We draw on these traditions to design a system that supports the representation and interpretation of complex, evolving information environments.

\subsection{Situational Awareness}

Our work is grounded in the situational awareness literature, particularly the three-level model of situational awareness proposed by Endsley \cite{endsley1988design}. In this framework, Level~1 situational awareness involves the perception of elements in the environment, including their status, attributes, and dynamics. Level~2 situational awareness concerns the comprehension of these elements and their significance within a broader context. Level~3 situational awareness involves the projection of future states based on an understanding of current conditions and trends \cite{endsley1988design}. Endsley defines situational awareness as ``the perception of the elements in the environment within a volume of time and space, the comprehension of their meaning, and the projection of their status in the near future'' \cite{endsley1988design}. This model has been widely used to reason about how information processing supports decision making in complex, time-sensitive environments.

We position our system as supporting the first two levels of situational awareness by improving perception and comprehension of the information environment. By structuring large-scale, dynamic social media data into temporally evolving narrative representations, the system provides a mechanism for filtering, organizing, and summarizing salient information. This supports information processing by reducing cognitive load and enabling analysts to observe how narratives emerge, persist, and change over time, rather than requiring direct engagement with raw data streams.

Our methodology is designed to support perception and comprehension through high-volume information pre-processing and prioritization via temporal windows and persistent cluster identity. Generative summarization is used as an abstraction mechanism to support interpretation of salient discourse, helping mitigate disruptions to information pre-processing, prioritization, confidence assessment, and interpretation that are known to degrade situational awareness and decision performance \cite{endsley2001disruptions}. Our approach also avoids the trap of supervised-based systems in having to pre-define topics, by instead surfacing emergent clusters in a way that enables human sense-making. Our approach addresses these challenges by providing a structured, temporally grounded representation of discourse without assuming predefined categories or static interpretations.

\subsection{Natural Language Tools Applied to Social Media Data}
A wide range of natural language processing techniques have been applied to social data to support large-scale analysis and monitoring. These approaches generally fall into three categories: classification-based methods, network-based analyses augmented with textual features, and temporal approaches designed to capture change over time. While each category offers useful capabilities, they also exhibit limitations when applied to dynamic information environments where narratives emerge, evolve, and dissipate among large volumes of data.

\subsubsection{Classification}

Numerous approaches apply large language models and fine-tuned neural models to classify social media data, including tasks such as stance detection, misinformation identification, and topic classification \cite{farr2024ens, ziems2024can}. These methods span supervised and weakly supervised pipelines as well as zero-shot and few-shot large language models \cite{brown2020language}. When label spaces are well defined and stable over time, classification-based approaches can efficiently characterize large volumes of data.

However, such approaches typically rely on predefined labels or coarse-grained categories (e.g., \emph{for}, \emph{against}, \emph{neutral}), which can obscure important nuance in complex information environments \cite{stancedef}. In rapidly evolving settings, narratives often emerge and transform in ways that are difficult to anticipate a priori, making static label sets challenging to define and maintain \cite{gama2014survey}. Moreover, classification pipelines generally operate on individual data points or snapshots in time, providing limited visibility into the changes in discourse over time or narrative lifecycles.

\subsubsection{Network-Based Methods}

Network-based approaches to social media analysis model information environments through the relational structure of user interactions, commonly using retweet, reply, or co-engagement graphs to identify communities, map information flows, and detect coordinated behavior \cite{starbird2019disinformation, beers2023followback}. These methods are particularly well-suited to identifying structural properties of discourse, such as community polarization, influence hierarchies, and cross-platform diffusion \cite{ng2022crossplatform}. By representing the information environment as a graph in which nodes are actors and edges are interactions, network-based analyses can reveal who amplifies what content and how narratives propagate through connected populations.

Despite these strengths, network-based approaches face inherent limitations in dynamic monitoring settings. Because interaction structure is often the primary unit of analysis, network methods tend to provide limited visibility into how the semantic content of narratives itself evolves over time. Retweet graphs can reveal which accounts amplify content, but provide limited insight into how the thematic substance of that content shifts in response to unfolding events. As discourse accelerates and narrative frames transform, network representations may lag or become structurally inconsistent, making it difficult for analysts to track narrative evolution as a continuous semantic process.

\subsubsection{Temporal Approaches}

A complementary tradition of work seeks to model discourse as a time-varying phenomenon, capturing how topics emerge, evolve, and dissipate over time. Dynamic topic models (DTMs), introduced by Blei and Lafferty \cite{blei2006dynamic}, extend latent Dirichlet allocation to track how the word distributions of topics shift across discrete time periods. These models have been widely applied to social media data to characterize temporal dynamics in online discourse, including event-driven shifts in public attention and the evolution of issue salience during crises \cite{churchill2022dynamic}.

More recent approaches leverage dense sentence embeddings and neural topic models to improve topic coherence and scalability on short, noisy social media text. Grootendorst's BERTopic \cite{grootendorst2022bertopic} clusters document embeddings in a shared semantic space and supports dynamic extensions that track topic evolution over time, enabling more fine-grained temporal tracking than traditional bag-of-words methods. Relatedly, Churchill and Singh \cite{churchill2022dynamic} propose topic-noise models specifically designed for social media data, incorporating noise distributions alongside topic distributions to separate coherent discourse from fragmented content.

While these temporal methods provide useful tools for characterizing how topics shift over time, they often rely on fixed topic counts, require periodic re-fitting, or produce representations that lack persistent identity across time windows. Topics extracted at one temporal interval are not inherently linked to semantically related topics at a previous interval, making it difficult to track a given narrative as a continuous evolving entity. Our approach addresses this limitation by maintaining explicit cluster identities across rolling windows, anchoring narrative structures in a stable embedding space that supports direct comparison of semantic states across time.

\section{System Overview}
Our system is designed to operate under common real-world constraints, including novel and dynamic data, discourse drift, and the absence of predefined labels, while explicitly modeling narrative emergence and change over time. Rolling temporal windows are used to allow topically relevant content to persist within narrative clusters while enabling those clusters to evolve as new data arrive. This design supports tracking how discourse within a given narrative shifts over time while handling narrative drift. An overview of our system is shown in Figure \ref{sys_over}.

To generate interpretable representations while controlling computational cost, the system selects a small set of representative posts from each cluster. Representative posts are chosen based on proximity to the centroid of the cluster in the embedding space, reflecting the central semantic structure of the cluster. To reduce redundancy, representatives are required to be sufficiently distinct from each other according to pairwise semantic similarity thresholds. These representative posts are then batched and passed to a generative language model to produce cluster-level summaries and broad thematic labels. The system is designed to operate continuously, updating narrative representations as new content arrives. In this way, these structures enable situational awareness for analysts that other, more constrained (e.g., supervised models) or less controlled (e.g.n simply exposing the cluster structure) systems do not.

\section{Methods}
\subsection{Data Collection}

Our dataset involves discourse surrounding the U.S. capture of Nicolás Maduro on January 3, 2026. In total, our dataset consisted of 9,914 X posts from 3,799 unique users between January 2 and January 7, 2026. We included posts with the keywords "maduro" or "venezuela" to ensure as broad coverage as possible. Additionally, due to data collection limits, we limited our collection to focus on posts that represented the most prominent discourse, using high re-post count as a proxy. Specifically, we sampled posts with a minimum of 50 re-posts, ensuring that any post included in our dataset had at least some traction within the broader online discussions making sense of the Maduro capture.

\subsection{Semantic Representation and Similarity}
Each collected post is embedded using a sentence-level transformer model (\texttt{all-mpnet-base-v2}). These embeddings are preserved throughout the pipeline, providing a shared semantic space in which both existing and newly arriving data points can be compared. Maintaining a consistent embedding space enables similarity-based operations across time, supporting temporal clustering and allowing narrative structure to evolve as new content emerges. Cosine similarity between embeddings is used as the primary measure of semantic relatedness throughout the system.

\subsection{Temporal Clustering}

Temporal clustering is central to the proposed methodology. Clusters are maintained across rolling temporal windows, allowing narrative structure to persist while incorporating newly arriving data. As the temporal window advances, new posts are embedded and assigned to existing clusters based on semantic similarity, and cluster centroids are recalculated to reflect updated content. Shifts in cluster centroids over time are used to capture narrative drift within the shared semantic space.

New clusters are created when incoming posts form coherent groupings in the embedding space that do not sufficiently match existing clusters, reflecting the emergence of new narratives. Conversely, clusters that do not receive assignments in a subsequent temporal window are not carried forward and are treated as having dissipated, reflecting narratives that no longer appear in the discourse. We use HDBSCAN as the clustering algorithm to avoid forcing cluster assignments and to allow semantically unrelated or sparse content to be treated as noise. 

Temporal window parameters are selected based on event dynamics in the applied setting. Given the rapid changes observed in the information environment of our case study, we use a twelve-hour temporal window that advances in four-hour increments. Global clusters are carried over into the next window if the cluster centroid has a cosine similarity greater than .85 with the cluster centroid in the previous window, meaning the new posts exist a similar shared semantic space while allowing for some shift in narrative.

\subsection{Representative Post Selection}

Representative posts are selected based on proximity to the cluster centroid in embedding space, reflecting the central semantic content of each cluster. To reduce redundancy and avoid selecting semantically equivalent posts (e.g., reposts or closely paraphrased content), a diversity constraint is applied by enforcing a minimum pairwise semantic distance among selected representatives. This approach balances semantic centrality with limited diversity, allowing representative posts to capture core cluster content while controlling input volume for downstream summarization.

Within each temporal window, a fixed number of representative posts (five) are displayed per cluster, while up to twenty representative posts are passed to the generative summarization and labeling stage. These values are configurable and may be adjusted to accommodate different use cases, data volumes, or computational constraints.

\subsection{Generative Summarization and Labeling}

At each temporal window, representative posts from each cluster are passed to a generative language model to produce a cluster-level summary and a broad thematic label. These summaries provide a compact abstraction of cluster content, while labels serve as lightweight descriptors that facilitate tracking narrative structure over time. Generative outputs are treated as interpretive representations of cluster discourse rather than definitive classifications, enabling rapid synthesis of evolving narratives within the information environment.

In the given case study, we use a commercial generative model API (GPT-4.1-mini); however, the summarization component is designed to be model-agnostic and can be replaced with either hosted or locally deployed generative models depending on operational constraints.

\section{System Implementation and Results}

To evaluate system behavior in a dynamic information environment, we applied the framework to social media discourse surrounding a U.S.-led security operation involving Venezuela in January 2026. System performance is evaluated through stratified human annotation of cluster assignments and analysis of temporal dynamics including centroid drift, lifecycle duration, and structural noise.

\subsection{Cluster Theme Performance}

To evaluate cluster theme performance, we curated a hand-labeled subset of 1{,}033 posts sampled across clusters and temporal windows (85 clusters total). Posts were selected using a stratified sampling strategy designed to capture variation within each cluster. For each cluster, posts were ordered by distance to the cluster centroid in embedding space. Five posts were randomly sampled from the closest region to the centroid (defined as the 25\% of posts with smallest centroid distance), five from a middle region (25\%–75\%), and five from the cluster periphery (75\%–100\%). For clusters containing fewer than 15 posts, all available posts were included. This stratified evaluation demonstrates that semantic proximity to cluster centroids provides a meaningful signal for both cluster interpretability and confidence estimation in dynamic information environments.

Using these annotations, we find that 91\% of posts are in an appropriate cluster. To better understand factors associated with correct cluster assignment, we conducted an analysis of variance (ANOVA) using centroid distance and cluster size as explanatory variables. Results indicate that centroid distance is a significant predictor of correct cluster classification, with posts closer to the centroid more likely to align with the dominant cluster theme. Cluster size also exhibited a measurable effect, suggesting that larger clusters tend to exhibit greater internal thematic consistency. Together, these findings provide an interpretable uncertainty signal at the individual post level and support the use of centroid proximity as a proxy for cluster confidence.

\subsection{Cluster Drift and Lifecycle}

Cluster drift is computed at the cluster level for narratives whose identity persists across consecutive temporal windows. Cluster identity is maintained when centroid cosine similarity exceeds 0.85 between adjacent windows; drift is then measured as the cosine distance between linked centroids. As a result, narrative drift reflects semantic change conditional on narrative persistence.

Across 75 narrative transitions (characterized by semantic shift across time windows), we observe substantial heterogeneity in semantic drift over time. The median cosine drift between consecutive temporal windows was 0.32, with a mean of 0.26 (SD = 0.20), indicating a right-skewed distribution of narrative change. Approximately 25\% of transitions exhibited minimal drift ($\leq$0.02), corresponding to stable narratives that persisted with little semantic change. In contrast, the upper quartile of transitions showed pronounced movement ($\geq$ 0.43), with the top decile exceeding 0.50 cosine distance, reflecting periods of rapid narrative restructuring. Together, these results indicate that the information environment simultaneously supports stable, enduring narratives and punctuated episodes of significant narrative evolution.

Narrative lifecycle analysis further reveals a dynamic information environment characterized by rapid turnover alongside a small number of persistent storylines. Across 37 detected narratives, the median consecutive persistence was two temporal windows (8 hours), indicating that most narratives were short-lived. However, a small subset of narratives exhibited substantially longer lifetimes, with the longest persisting for 22 consecutive windows (88 hours). This long-lived narrative corresponded to mainstream reporting of the U.S. operation in Venezuela and functioned as the dominant narrative frame throughout the event. The resulting lifecycle distribution reflects a system composed of transient components operating alongside a limited number of stable structures, consistent with expectations for complex, non-stationary information environments.

\begin{figure*}[h] 
    \centering
    \begin{subfigure}[b]{0.45\textwidth}
        \centering
        \includegraphics[width=\textwidth]{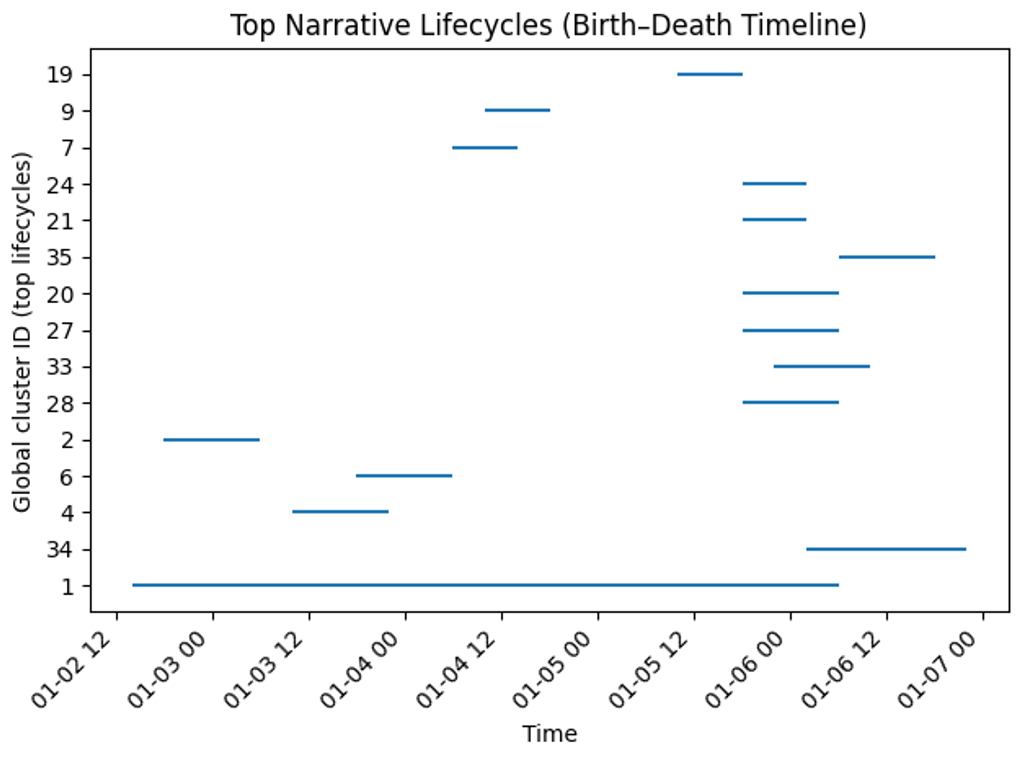}
        \caption{ Distribution of narrative lifecycles, showing a small number of long-lived narratives alongside a larger set of short-lived, transient narratives. }
        \label{lifecycle}
    \end{subfigure}
    \hfill
    \begin{subfigure}[b]{0.45\textwidth}
        \centering
        \includegraphics[width=\textwidth]{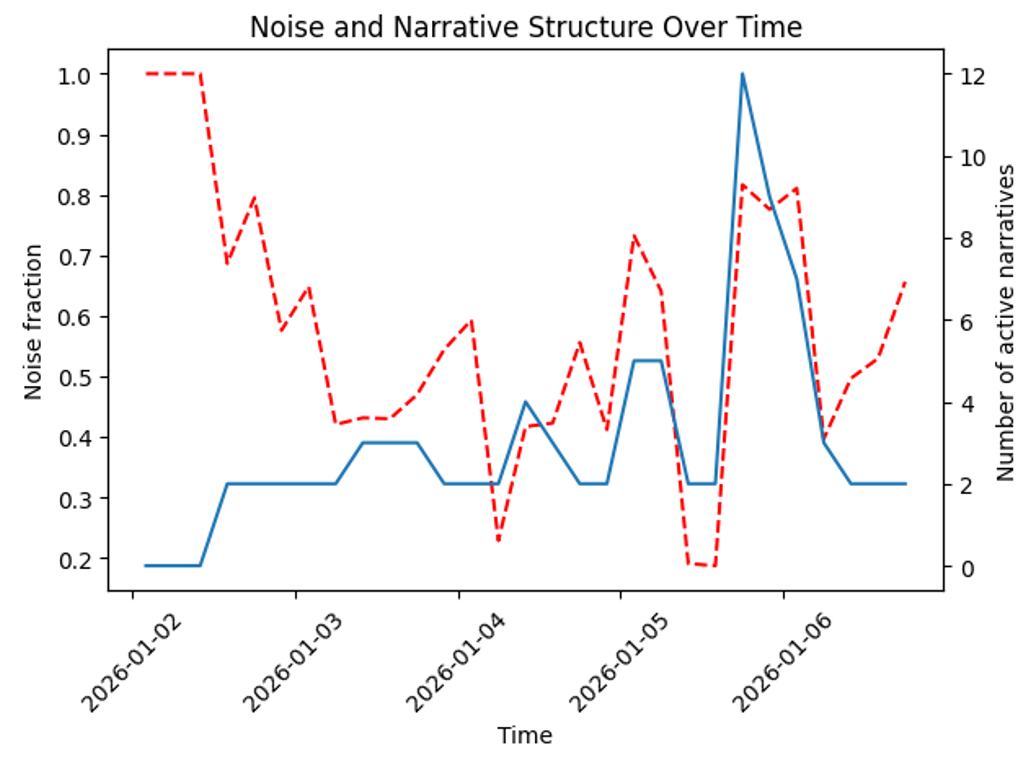}
        \caption{Proportion of posts classified as noise and number of detected clusters across temporal windows, illustrating periods of narrative emergence, consolidation, and disruption.}
        \label{noise}
    \end{subfigure}
\end{figure*}
\subsection{Cluster Structure and Noise Over Time}

To characterize how narrative structure evolves over the course of the event, we examine the proportion of posts classified as noise alongside the number of detected clusters within each temporal window. Noise is defined as content that does not meet density requirements for inclusion in a coherent semantic cluster and is not forced into an existing narrative structure.

Figure \ref{noise} shows the fraction of noise posts and the number of detected clusters over time. Early in the event, a relatively high proportion of content is classified as noise, reflecting fragmented and weakly structured discourse as narratives begin to form. As the event progresses, the number of clusters increases while the noise fraction declines, indicating consolidation of discourse into coherent narrative structures. Periods of rising noise correspond to moments of narrative disruption or rapid discourse expansion, where new or weakly defined themes emerge before stabilizing.

Across the full event window, the mean noise fraction was \textbf{0.49}, with values ranging from \textbf{.19} to complete noise during initial data collection. The observed inverse relationship between noise proportion and cluster count suggests that the system dynamically adapts to changes in discourse structure, allowing narratives to emerge, fragment, and recombine without enforcing artificial coherence.

\section{Discussion}

The observed results demonstrate that narrative structure within the information environment can be modeled as an evolving system state governed by measurable semantic dynamics. Density-based clustering combined with temporal linkage enables persistent narrative identity while capturing controlled semantic drift. Observed lifecycle variation indicates that information environments simultaneously contain transient discourse fragments and stable narrative anchors.

These findings demonstrate that narrative dynamics can be represented as measurable state transitions within a shared semantic space, supporting scalable monitoring without predefined taxonomies.

\subsection{Implications for Cognitive Load Management}

Our methodological framework reduces cognitive burden by decomposing a dynamic and complex environment into interpretable narrative clusters that evolve with the event. Approximately half of collected content was filtered as noise, while remaining posts were organized into 37 global narratives spanning the event. Temporal continuity stabilizes interpretation by allowing individuals to track evolving narratives rather than repeatedly reconstruct categories. Centroid proximity provides an interpretable confidence signal that supports prioritization under time constraints.

\section{Conclusion}

We present a systems-oriented framework for understanding dynamic information environments as temporally evolving semantic structures. Using a situational awareness theory framework, we  transform large-scale social media streams into coherent narrative clusters that enable monitoring without predefined labels. Application to a real-world crisis demonstrates that narrative emergence, persistence, and drift can be quantitatively tracked, supporting decision-making in dynamic information environments.

\subsection{Limitations}

As with any system designed to support situational awareness in complex environments, this approach has several limitations. First, temporal window sizes and similarity thresholds must be specified in advance. Although selected to reflect event dynamics in the present case study, alternative parameterizations may yield different narrative lifecycles and drift patterns.

Second, the system prioritizes a high-level structural view of the information environment. While this enables large-scale monitoring and temporal tracking, it may overlook granular narrative nuances that qualitative or micro-level analyses can reveal. There is an inherent trade-off between scalable structural modeling and fine-grained interpretive depth.

Third, like any system based on unsupervised machine learning, this system is meant to augment human situational awareness. Clusters from unsupervised learning require interpretation to determine their meaning and value. This system design is not meant to be a stand-alone system, but one which augments an analyst's situational awareness.

Finally, the framework relies on multiple modular components, including embedding models, clustering algorithms, and generative summarization systems. Although this modularity allows flexibility and substitution of components, it also introduces computational overhead and potential sensitivity to upstream modeling choices.

\subsection{Acknowledgments}

Generative AI tools were used for drafting assistance and software development support. All analysis and results were verified by the authors. The full implementation is available at https://github.com/davidthfarr/sensemaking\_project, and an interactive dashboard is hosted at https://davidthfarr.github.io/sensemaking/.

%Bibliography 
\bibliographystyle{plain}
\bibliography{sample}

\end{document}